\documentclass[sigconf]{aamas}

\usepackage{balance} 
\usepackage{algorithm}
\usepackage{algorithmic}
\usepackage{amsmath}
\usepackage{graphicx}
\usepackage{amsmath}
\usepackage{diagbox}
\usepackage{slashbox}
\usepackage{color}
\usepackage{mathtools}
\usepackage{float}
\usepackage{multirow}
\usepackage{caption} 

\doi{RNLD7242}

\makeatletter
\gdef\@copyrightpermission{
  \begin{minipage}{0.2\columnwidth}
   \href{https://creativecommons.org/licenses/by/4.0/}{\includegraphics[width=0.90\textwidth]{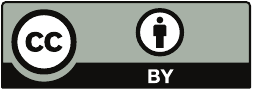}}
  \end{minipage}\hfill
  \begin{minipage}{0.8\columnwidth}
   \href{https://creativecommons.org/licenses/by/4.0/}{This work is licensed under a Creative Commons Attribution International 4.0 License.}
  \end{minipage}
  \vspace{5pt}
}
\makeatother

\setcopyright{ifaamas}
\acmConference[AAMAS '26]{Proc.\@ of the 25th International Conference
on Autonomous Agents and Multiagent Systems (AAMAS 2026)}{May 25 -- 29, 2026}
{Paphos, Cyprus}{C.~Amato, L.~Dennis, V.~Mascardi, J.~Thangarajah (eds.)}
\copyrightyear{2026}
\acmYear{2026}
\acmDOI{}
\acmPrice{}
\acmISBN{}

\title[AAMAS-2026 Formatting Instructions]{Rethinking Priority Scheduling for Sequential Multi-Agent Decision Making in Stackelberg Games}

\author{Xiangyu Liu}
\affiliation{
  \institution{Dalian University of Technology}
  \city{Dalian}
  \country{China}}
\email{2728320168@mail.dlut.edu.cn}

\author{Liang Zhang}
\affiliation{
  \institution{Dalian University of Technology}
  \city{Dalian}
  \country{China}}
\email{liangzhang@dlut.edu.cn}

\author{Bo Jin}
\affiliation{
  \institution{Dalian University of Technology}
  \city{Dalian}
  \country{China}}
\email{jinbo@dlut.edu.cn}

\author{Ziqi Wei}
\authornote{Corresponding author}
\affiliation{
  \institution{Dalian University of Technology}
  \city{Dalian}
  \country{China}}
\email{ziqi.wei@ualberta.ca}

\begin{abstract}
Current research applying N-level Stackelberg Game to multi-agent systems often uses the default decision order of agents provided by the environment. However, this raises the question: \textbf{Does the order of agents necessarily affect the final equilibrium point of the game?} To address this, we formally analyze the N-level Stackelberg Game in which changing the order where the agents make decisions typically leads to an overdetermined system. As a result, the equilibrium point is shifted unless special structural conditions are met. Based on this, we propose the Hierarchical Priority Adjustment (HPA) method, which adjusts and selects the agents’ decision order. For the upper level, an upper policy dynamically selects the optimal decision order of agents based on the current game state; for the lower level, agents execute the strategy in the Spatio-Temporal Sequential Markov Game (STMG) based on the selected order. To coordinate learning across time scales, we employ a slow-fast update scheme with shared intrinsic rewards derived from the upper policy advantage function. Experimental results on high-precision control tasks such as multi-agent MuJoCo show that HPA outperforms the benchmark algorithms and robustly adapts to changing environments. These results highlight the crucial role of optimizing the decision order of agents in N-level Stackelberg Game. 
\end{abstract}


\newcommand{\BibTeX}{\rm B\kern-.05em{\sc i\kern-.025em b}\kern-.08em\TeX}

\begin{document}


\pagestyle{fancy}
\fancyhead{}


\maketitle 


\section{Introduction}

In Multi-Agent Systems (MAS), agents independently complete their own tasks to achieve individual optimization. In addition, they also cooperate with each other to achieve shared global goals. Multi-Agent Reinforcement Learning (MARL) has become a promising approach to solve this task, nevertheless it brings challenges. In a constantly changing environment, the rewards of agents are usually closely related to the actions of others. To achieve optimal and stable collaborative behavior, each individual in the system needs to form a consistent and stable collaborative strategy through continuous interaction. This requires the construction of an effective and efficient interaction structure between agents.

Game theory provides an effective conceptual framework for promoting the interaction among agents, which offers a promising approach to solving multi-agent collaboration. Initially, some studies model multi-agent systems by using the concept of Nash Equilibrium (NE) in non-cooperative games, expecting that collaborative strategies can converge to NE point, such as Nash Q-Learning \cite{r1}, Mean Field Q-learning \cite{r2} and HATRPO \cite{r3}. However, these methods lead to multiple NE points existing in the game, and in this case, agents choose different NE strategies so that the overall strategy can't converge to the one NE point. For instance, as can be observed in the left panel of Figure 1, even within relatively simple game environments, multiple NE points may exist. When agents make decisions simultaneously, this can lead to different agents optimising towards divergent directions, making it difficult for their strategies to converge.

Therefore, subsequent research turned to find the  Stackelberg Equilibrium (SE) \cite{r4}, where agents adopt asynchronous execution strategy and make decisions in a leader-follower framework. Leaders prioritize decision-making and implement their policies, while followers respond rationally to such implementation based on the leaders' decisions. This approach mitigates environmental non-stationary issues by treating the remaining followers as a part of the environment when the leader acts. Moreover, the SE point demonstrates superior coordination performance compared to the NE point. In Figure 1, regardless of whether the environment is set to a global objective or self-reward, the payoff at the SE point is higher than that at the NE point. This arises from the continuous accumulation of Pareto advantages through asynchronous decision-making between characters during the Stackelberg game. Early research divided agents into leader and follower (2-level Stackelberg Game) to establish an asynchronous interaction structure, such as Bi-AC \cite{r5} and ST-MADDPG \cite{r6}, but this method still has the same issues as the above method for finding NE points. Although it achieves asynchronous execution between the two levels, the agents within the same level are still executed synchronously.

\begin{figure}[ht]
    \centering
    \setlength{\tabcolsep}{3pt}
    \begin{minipage}[b]{0.2\textwidth}
        \centering
        \begin{tabular}{|c|c|c|c|}
            \hline
            \diagbox{{\(a_{1}\)}}{{\(a_{2}\)}}& \(a_1^2\) & \(a_2^2\) & \(a_3^2\) \\
            \hline
            \(a_1^1\) & -10& 0 & \textcolor{red}{10} \\
            \hline
            \(a_2^1\) & 0 & \textcolor{blue}{2} & 0 \\
            \hline
            \(a_3^1\) & \textcolor{blue}{8} & 0 & -10\\
            \hline
        \end{tabular}
    \end{minipage}
    \hfill
    \begin{minipage}[b]{0.25\textwidth}
        \centering
        \begin{tabular}{|c|c|c|c|}
            \hline
            \diagbox{{\(a_{1}\)}}{{\(a_{2}\)}} & \(a_1^2\) & \(a_2^2\) & \(a_3^2\) \\
            \hline
            \(a_1^1\) & \textcolor{red}{0,5} & -10, -5 & -8, 4 \\
            \hline
            \(a_2^1\) & -5, -10 & \textcolor{blue}{-5,0} & -15, -5 \\
            \hline
            \(a_3^1\) & 5, 0 & -10, -5 & -10, 5 \\
            \hline
        \end{tabular}
    \end{minipage}
    
    \caption{Matrix games. Left:  \((a_1^1, a_3^2)\), \((a_2^1, a_2^2)\) and \((a_3^1, a_1^2)\) are NE points, and \((a_1^1, a_3^2)\) is the only SE point. Right: The payoff matrix of the Mixing game. When \(a_{1}\) is the leader, it has only one NE point \((a_2^1, a_2^2)\) and only one SE point \((a_1^1, a_1^2)\). The SE is Pareto superior to the NE.}
    \label{fig:matrix_games}
\end{figure}
To solve this issue, some research extended 2-level Stackelberg Game to N-level Stackelberg Game. These approaches serialized the execution of multi-agent systems, forming Stackelberg sub-games level by level through sequential execution, and continuously accumulating Pareto advantages. It can continuously enhance the level of collaboration among multi-agent. STEP \cite{r7} proposed Spatio-Temporal Sequential Markov Game (STMG), providing a complete process for modeling multi-agent systems with N-level Stackelberg Game and achieving good performance. STEER \cite{r9} combined Stackelberg Game (SG) with Transformer, fully leveraging the advantages of both, and achieved excellent experimental results.

The N-level Stackelberg Game essentially serializes agents in the system, creating differences in the order of execution between agents. The Stackelberg game originates as an economic concept, used to describe the strategic interactions between firms of differing scale and status within a market. It enables the determination of whether a firm acts as a leader or follower based on quantifiable metrics. However, when applied to multi-agent system, it is often impractical to accurately quantify an agent's capabilities, thereby precluding the definitive classification of an agent's character within the environment. In summary, determining an agent's identity remains challenging.

Recently, some research begun to focus on the issue of agent execution sequence. The B\&P \cite{r10} algorithm determined who plays first through a heuristic approach, and solved the multi-agent collaborative navigation problem by combining real physical systems. Combining the research content of this paper, we rethink SG and MAS and propose the following two research questions: \textbf{(a) Does the execution sequence of agents necessarily have an impact on the N-level Stackelberg Game? \textbf{(b)} If so, how should the execution sequence of agents be determined?} Although the B\&P algorithm implemented dynamic adjustment of agent sequences, it didn't provide theoretical guarantees that the change of agent sequences will affect the final equilibrium point of the game. In addition, its efficiency and performance depend on the sub-problem solver and pruning strategy, which all need to be designed manually according to the application domain, so it lacks universality. Moreover, when facing high-dimensional continuous state action spaces and nonlinear dynamic tasks in MARL, it is more likely to learn suboptimal results compared to deep learning models. We aim to propose a universal end-to-end method that can learn the optimal execution sequence in complex MARL environment tasks for different situations. 

In this paper, we prove the question \textbf{(a)} theoretically that different execution sequences significantly affect the equilibrium of the N-level Stackelberg Game by analyzing the concept and mechanism of action of Stackelberg Game. Based on this, the HPA algorithm is proposed to learn the optimal decision sequence of multiple agents in different situations to solve the above questions \textbf{(b)}. The contributions are summarized as follows:
\begin{enumerate}
    \item From the perspective of game theory, we use mathematical methods to prove that in the same situation, the execution sequence will significantly affect the equilibrium of the game, and thereby influence the collaboration level of multiple agents.
    \item We propose the HPA algorithm. In a complex multi-agent interaction environment, as the game progresses, it can select the best execution sequence in different situations, thereby enhancing the overall performance. The results in more complicated scenarios also illustrate its superiority over powerful benchmarks in terms of the overall performance. 
\end{enumerate}

\section{The Preamble and Related Work}

\subsection{Related Work}
In some studies, Stackelberg Equilibrium (SE) is adopted as the convergence objective for multi-agent systems, where leaders can access followers’ reward information and maintain copies of their value functions. Bi-AC \cite{r5} introduces a bi-level actor–critic framework with a Q-learning-based leader and DDPG-based followers, requiring both parties to retain key networks during execution. Similarly, ST-MADDPG \cite{r6} follows the CTDE paradigm by dividing agents into leader and follower levels, with followers observing leaders’ actions. However, these approaches are limited to 2-level structures, and extending them to multiple levels is non-trivial. Subsequent work generalised two-level Stackelberg games to N-level settings. For example, STEP \cite{r7} uses a conditional hypernetwork to construct agent policies, reducing model complexity while enabling parallel execution, whereas STEER \cite{r9} leverages the autoregressive structure of Transformers to model sequential execution and integrates global information through unified encoding.

During experimentation, the ordering of agents naturally becomes a critical issue that prior work largely overlooks. B\&P \cite{r10} addressed this by dynamically adjusting agent sequences using heuristic rules, but this approach depends on domain knowledge and manual pruning, which limits scalability. Therefore, effectively resolving agent serialisation remains an important challenge in multi-agent systems.

\subsection{Spatio-Temporal Sequential Markov Game}
The STMG is an evolutionary version of Markov Game (MG). It is defined as the tuple
\( \Gamma =\left\langle\mathcal{I}, \mathcal{S},\left\{\mathcal{A}^{i}\right\}_{i \in \mathcal{I}}, \mathcal{P},
\left\{r^{i}\right\}_{i \in \mathcal{I}}, \gamma,\left\{h^{i}\right\}_{i \in \mathcal{I}}\right\rangle\), where \(\mathcal{I}\) represents the set of all agents with \(|\mathcal{I}|=n\), and \(s \in \mathcal{S}\) represents the environmental state. \(a^{i} \in \mathcal{A}^{i}\) is the action of agent \(i\) and the joint action space is \(\mathcal{A}=\prod_{i=1}^{n} \mathcal{A}^{i}\). \(\mathcal{P}: \mathcal{S} \times \mathcal{A} \rightarrow \Omega(\mathcal{S})\) represents the state transition function of the environment, where \(\Omega(X)\) denotes the set of probability distributions over \(X\). \(r^{i}: \mathcal{S} \times \mathcal{A} \rightarrow \mathbb{R}\) is the reward function of agent \(i\) and \(\gamma\) is the discount factor. \(h^{i}\) denotes the decision priority of agent \(i\) and \(\mathcal{H}=\left\{h^{1}, \ldots, h^{n}\right\}\) is a prioritized permutation of agents. 

At time \(t\) step , the agent with priority \(h^{i}\) executes its strategy \(\pi^{h^{i}}: \mathcal{S} \times \mathcal{A}^{h^{1}} \times \cdots \times \mathcal{A}^{h^{i-1}} \rightarrow \Omega\left(\mathcal{A}^{h^{i}}\right)\) based on the subgame state \(s_{t}^{h^{i}}=\left(s_{t}, a_{t}^{h^{1}}, \ldots, a_{t}^{h^{i-1}}\right)\). The environment transitions to a new state \(s_{t+1} \sim P\left(s_{t+1} \mid s_{t}, \boldsymbol{a}_{\boldsymbol{t}}\right)\) after receiving the joint action \(\boldsymbol{a}_{\boldsymbol{t}}=\left(a_{t}^{1}, \ldots, a_{t}^{n}\right)\) and assigns private rewards \(r^{i}\left(s_{t}, \boldsymbol{a}_{\boldsymbol{t}}\right)\) for each agent. The joint policy is represented by \(\boldsymbol{\pi}\left(s_{t}\right)=\prod_{i=1}^{n} \pi^{h^{i}}\left(s_{t}^{h^{i}}\right)\). The transition function and the joint strategy determine the marginal distribution of the state at each time step, i.e., \(s \sim \rho_{\pi}\). Within this framework, each agent aims to maximize its own discounted cumulative reward \(R^{i}(\tau)=\sum_{t=0}^{T} \gamma^{t} r^{i}\left(s_{t}, \boldsymbol{a}_{\boldsymbol{t}}\right)\) over a trajectory \(\tau\) of length \(T\). 

\subsection{Example of 2-Level Stackelberg Game}
Here we use a simple example to illustrate question \textbf{(a)}: the influence of the execution sequence on the equilibrium state of the game. In the SG scenario, there are two characters \(a_{1}\) and \(a_{2}\), and the payoff matrix is shown in Table 1. When \(a_{1}\) is the leader and \(a_{2}\) is the follower, \(a_{2}\) chooses the action that maximizes its own benefit under the condition that \(a_{1}\) has made decision. The benefits of the actions are (40,40) and (20,20) respectively, and \(a_{1}\) chooses the action that maximizes its own benefit among these actions, i.e. \((a_{1}^{1},a_{2}^{1})\) is the SE point. Similarly, when \(a_{2}\) is the leader, (\(a_{1}^{2}\),\(a_{2}^{2}\)) is the SE point.
\begin{figure}
    \centering
    \begin{tabular}{|c|c|c|}\hline
         \diagbox{\(a_{1}\)}{\(a_{2}\)}& \(a_{2}^{1}\)& \(a_{2}^{2}\)\\\hline
         \rule{0pt}{3ex}\(a_{1}^{1}\) & \textcolor{red}{40,40} & 0,0 \\[1ex]\hline
         \rule{0pt}{3ex}$a_{1}^{2}$ & 80,0 & \textcolor{red}{20,20} \\[1ex]\hline
    \end{tabular}
    \caption{The example payoff matrix of 2-level Stackelberg Game.}
    \label{tab:my_table}
\end{figure}
It can be found that the sequence of role decisions affects the final equilibrium result of the game. It also prompts us to consider whether the suboptimal performance of the model proposed in previous research in certain test environments can be attributed to the predetermined decision sequence, rather than reflecting the actual performance limitations of the model itself.

\subsection{N-Level Stackelberg Game}
SG is a well-established game-theoretic framework that models hierarchical decision-making structures where some agents have advantages over others. Typically, such structures consist of leaders, who are superior agents capable of committing to their actions prior to other agents; and followers, who are inferior agents that must respond to the leaders’ decisions.In the context of multi-agent Stackelberg games, each agent is placed in a different priority position to match the decision-making mode of STMG. This process is extended from the 2-level Stackelberg Game structure to the N-level Stackelberg Game. And it gives rise to an N-level optimization problem:
\begin{equation}
\begin{aligned}
&\max _{\pi^{i} \in \Pi^{i}} \mathcal{J}^{i}\left(\pi^{1: i-1}, \pi^{i}\right) 
,  \pi^{j} \in \arg \max _{\pi^{j^{\prime}} \in \Pi^{j}} 
    \mathcal{J}^{j}\left(\pi^{1: j^{\prime}-1}, \pi^{j^{\prime}}\right)\\
&\max _{\pi^{j} \in \Pi^{j}} \mathcal{J}^{j}\left(\pi^{1: j-1}, \pi^{j}\right)
\end{aligned}
\end{equation}
where \(i \in[1: n]\) and \(j \in[i+1, n]\). Within the hierarchical decision-making structure of STMG, each agent assumes the role of a follower to higher-level agents while simultaneously acting as a leader to lower-level agents. For followers, they receive decision information from superior agents during both the execution and training procedures. The policy gradients of the agents are then updated in the direction of the optimal response to leaders, yielding an approximation of the solution to the inner optimization problem.On the other hand, for leaders, they interact with the environment and perceive the reaction of the inferior agents. When updating their policies, leaders consider followers as part of the surrounding environment and maximize their rewards, resulting in an approximate solution to the outer optimization problem.

Under the RL training paradigm, all agents possess the capability to maximize their utility in accordance with current conditions. Through continuous interaction with the environment, agents eventually arrive at a consensus wherein inferior agents execute optimal responses to the decisions of superior agents, and superior agents optimize their policies based on this premise, promoting the attainment of SE policies by all agents. 

\section{Priority is Important}

This section aims to solve the question \textbf{(a)}: \textbf{Does the execution sequence of agents necessarily have a impact on the N-level Stackelberg Game?} The question centers on whether a change in the order of execution of the N-level Stackelberg Game changes the final game payoff, i.e., the SE point changes. From the perspective of Stackelberg game, each character is inherently unequal in ability and status. Intuitively, characters with high ability and status should have high priority to achieve a better global SE, and vice versa.  

The proof is given below: first, the game system is setup as follows. There are \(n\) players in the game with corresponding strategies \(\pi_{i}\in \Pi_{i}\). The strategy space is also the action space, and here we specify the action space as continuous variables, such as the movement and rotational angles of a robot's joints. For discrete variables, we can derive them using the tabular method. Theoretically, results can be obtained even for highly complex game settings. The payoff function is defined as \(Q_{i}(\pi_{1},...,\pi_{n})\). The payoff function represents the correspondence between the actions taken by the players during the game and the payoffs they receive. Based on the properties of the Stackelberg game, the strategy \(\pi_{i}\) can be written as \(\pi_{i}=f(\pi_{1},...,\pi_{i-1})\).

Subsequently, the first-order conditions characterizing the optimal strategy of character \(i\) in the general case are derived through backward induction \cite{r19}. These conditions serve as necessary criteria for optimality in sequential decision-making:
\begin{itemize}
    \item Character \(n\) maximizes its own return \(Q_{n}\) given the decisions of  \(1,...,n-1\) characters \((\pi_{1}^{*},...,\pi_{n-1}^{*})\). The objective function can be formalized as:
    \begin{equation}
         {\arg\max} _{{\pi_{n}} \in {\Pi_{n}}} Q_{n}\left({\pi_{1}^{*}:\pi^{*}_{n-1}}; {\pi_{n}}\right)
    \end{equation}
    The condition obtained is:
    \begin{equation}
        \frac{d Q_{n}}{d \pi_{n}}=0
    \end{equation}

    \item Similarly, when character \(i\) is given the strategies of the former \(i-1\) characters \((\pi_{1}^{*},...,\pi_{i-1}^{*})\), it chooses \(\pi_{i}\) to maximize \(Q_{i}\). To make sure, the strategies of the latter characters are taken into consideration. The object function is formalized as:
    \begin{equation}
    {\arg \max} _{{\pi_{i}} \in {\Pi_{i}}} Q_{i}({\pi_{1}}^{*}:\pi_{i-1}^{*}; {\pi_{i}}; {\pi_{i+1}}:{\pi_{n}})
    \end{equation}
    The condition obtained is:
    \begin{equation}
        \frac{d Q_{i}}{d \pi_{i}}=\frac{\partial Q_{i}}{\partial \pi_{i}}+\sum_{j=i+1}^{n} \frac{\partial Q_{i}}{\partial \pi_{j}} \cdot \frac{d \pi_{j}}{d \pi_{i}}=0
    \end{equation}
    where \(\frac{d \pi_{j}}{d \pi_{i}}\) can be expanded by the chain rule as
    \begin{equation}
        \frac{d \pi_{j}}{d \pi_{i}}=\sum_{k=i}^{j-1} \frac{\partial \pi_{j}}{\partial \pi_{k}} \cdot \frac{d \pi_{k}}{d \pi_{i}}
    \end{equation}
    \item Character \(1\) maximizes its own return \(Q_{1}\). The objective function can be formalized as:
\begin{equation}
         {\arg\max} _{{\pi_{1}} \in {\Pi_{1}}} Q_{1}\left(\pi_{1};\pi_{2}: {\pi_{n}}\right)
    \end{equation}
    The condition obtained is:
    \begin{equation}
        \frac{d Q_{1}}{d \pi_{1}}=\frac{\partial Q_{1}}{\partial \pi_{1}}+\sum_{j=2}^{n} \frac{\partial Q_{1}}{\partial \pi_{j}} \cdot \frac{d \pi_{j}}{d \pi_{1}}=0
    \end{equation}
    where \(\frac{d \pi_{j}}{d \pi_{1}}\) can be expanded by the chain rule as
    \begin{equation}
        \frac{d \pi_{j}}{d \pi_{1}}=\sum_{k=1}^{j-1} \frac{\partial \pi_{j}}{\partial \pi_{k}} \cdot \frac{d \pi_{k}}{d \pi_{1}}
    \end{equation}
    \item Summarize: for any character \(i\), the necessary condition to be satisfied for a strategy \(\pi_{i}\) that maximizes \(Q_{i}\):
    \begin{equation}
    \frac{d Q_{i}}{d \pi_{i}}=\frac{\partial Q_{i}}{\partial \pi_{i}}+\sum_{j=i+1}^{n}\left(\frac{\partial Q_{i}}{\partial \pi_{j}} \cdot \sum_{k=i}^{j-1} \frac{\partial \pi_{j}}{\partial \pi_{k}} \cdot \frac{d \pi_{k}}{d \pi_{i}}\right)=0
    \end{equation}
\end{itemize}
Noted that the mathematical computation of the maximum value usually requires the second-order derivative condition to distinguish the category of extreme values, but here we decide not to compute the second-order derivative condition. The reason is that the premise of our derivation is the necessary conditions that need to be satisfied by the SE point, which means that the second-order derivative condition naturally holds. 

It can be found in Equation 9 that the characters other than \(1\) and \(n\) act as both the leader and the followers in the game, which means that their decisions are obtained by the simultaneous action of both parts. Corresponding to the two terms: for the first term \(\frac{\partial Q_{i}}{\partial \pi_{i}}\), the character \(i\) acts as a follower and chooses the strategy that maximizes \(Q_{i}\) when given the strategy \((\pi_{1}^{*},..,\pi_{i-1}^{*})\); For the second term \(\sum_{j=i+1}^{n}\left(\frac{\partial Q_{i}}{\partial \pi_{j}} \cdot \sum_{k=i}^{j-1} \frac{\partial \pi_{j}}{\partial \pi_{k}} \cdot \frac{d \pi_{k}}{d \pi_{i}}\right)\), it acts as a leader, which needs to forecast the decisions of subsequent characters to make decisions that are in one's self-interest, as reflected in the chain of derivatives.

Having obtained the first-order derivative condition satisfied by the optimal strategy of character \(i\), next we need to prove whether the formed game structure when the order of the characters changes still satisfies the conditions required before the order change. Next we will prove whether there exists an overall strategy that satisfies both sets of equations before and after the change of the characters order:
\begin{itemize}
    \item Original game structure \(\Gamma\): the order of characters is \(1,2,...,n\). The best strategy under this order is \(\pi^{*}=(\pi_{1}^{*},...,\pi_{n}^{*})\), \(\pi \in \Pi\). The payoff functions are \(Q_{i}=(\pi_{1},...,\pi_{n})\), respectively. The equilibrium condition for character \(i\) is as follows:
    \begin{equation}
    F_{i}(\pi_{i})=\frac{\partial Q_{i}}{\partial \pi_{i}}+\sum_{j=i+1}^{n} \frac{\partial Q_{i}}{\partial \pi_{j}} \cdot \frac{d \pi_{j}}{d \pi_{i}}=0
    \end{equation}

    \item Order-shuffled game structure \(\Gamma^{'}\): based on the order of characters in the original game, there are at least two roles that switch places with each other. The best strategy under this order is \(\pi^{*'}=(\pi_{1}^{*'},...,\pi_{n}^{*'})\), \(\pi' \in \Pi'\). The equilibrium condition for character \(i\) is as follows:
    \begin{equation}
   F_{i}'(\pi_{i})= \frac{\partial Q_{i}}{\partial \pi_{i}^{\prime}}+\sum_{j=i+1}^{n} \frac{\partial Q_{i}}{\partial \pi_{j}^{\prime}} \cdot \frac{d \pi_{j}^{\prime}}{d \pi_{i}^{\prime}}=0
    \end{equation}
\end{itemize}
Although the condition formulas for character i before and after the change in order remain the same, the hierarchical relationship between its preceding and following characters changes. We combine the two sets of equations by rows to form the set of equations \(J\):
\begin{equation}
    \begin{array}{c}
F(\pi)=\left(F_{1}(\pi_{1}); \ldots; F_{n}(\pi_{n})\right)=0 \\
F^{\prime}(\pi)=\left(F_{1}^{\prime}(\pi_{1}); \ldots; F_{n}^{\prime}(\pi_{n})\right)=0 \\
\end{array}
\end{equation}
\begin{equation}
    J(\pi)=(F(\pi);{F^{\prime}(\pi)})=0
\end{equation}
We express the set of equations \(J(\pi)=0\) in the form of \(A \pi=b\), where \(A \in \mathbb{R}^{2n\times n}\) is the coefficient matrix with respect to \(\pi=(\pi_{1},...,\pi_{n}), \pi \in \Pi \cap \Pi^{\prime}\). \(b\) is a vector composed of the opposite numbers of the constant terms in the set of equations, which is a non-zero vector. It is noted that matrix A is overdetermined and heterogeneous, so there is no strategies \(\pi\) that can simultaneously satisfy all constraints. A solution exists when the following conditions are met: 
\begin{enumerate}
    \item If it is a set of linear equations, there will only be a solution when \(r(A)=r(A|b)\) \cite{r11}.
    \item If it is a set of nonlinear equations, in mathematics, iterative methods such as LM \cite{r12} or Gauss-Newton \cite{r13} are usually used to solve them. The specific approach is as follows:
\end{enumerate}
\begin{equation}
E(\boldsymbol{\pi}) \stackrel{\text { def }}{=} \sum_{i=1}^{2n} F_{i}^{2}(\boldsymbol{\pi})=\|J(\boldsymbol{\pi})\|^{2}
\end{equation}\begin{equation}
    \boldsymbol{\pi}^{*}=\arg \min _{\boldsymbol{\pi}}\|J(\boldsymbol{\pi})\|^{2}
\end{equation}
Given an error coefficient \(\epsilon\), when the error \(\|J(\boldsymbol{\pi})\|^{2} < \epsilon\), it is considered that a solution exists. Finally, it can be concluded that, under normal circumstances, no single strategy exists that can simultaneously satisfy the constraints both before and after a change in order. 

\section{Priority Scheduling for SG}
The intuition behind the model: for instance, in a basketball game, during offensive plays, the actions of forwards are relatively more important than those of midfielders and defenders, whereas the opposite holds true during defensive plays. In the previous section, we prove that the adjustment of the order of the characters affects the final gain of the game in the same state. Based on this, from the perspective of multi-agent system, with the game process keep going, the state of the system keeps changing, and the optimal execution order of the agents will also change. We propose \textbf{H}ierarchical \textbf{P}riority \textbf{A}djustment to solve the question \textbf{(b)}: \textbf{how should the execution sequence of agents be determined?}

Our proposed method centers on choosing the optimal order of execution of the agents in the different states of the system. It can be naturally associated with hierarchical reinforcement learning (HRL), whose core idea lies in the decomposition of a complex task into more manageable sub-tasks. It's optimization and decision-making can be carried out separately between levels, which is a natural fit with our problem.
\begin{algorithm}[H]
\caption{Hierarchical Priority Adjustment (HPA)}
\begin{algorithmic}[1]
\STATE Initialize environment state $s_0$, pretrain lower-level policies $\pi_1, ..., \pi_n$, time interval $k$
\STATE Initialize upper policy $\pi_\Omega$, termination function $\beta_{\omega,\phi}$, critic $V_\Omega$
\STATE $s \gets s_0$
\STATE Sampling order $w$ from distributions generated by the strategy model $\pi_\Omega(s)$

\FOR{$episode =0$ to $episodes$}
    \FOR{$t = 0$ to $episode-length$}
        \STATE For each agent $i$ (according to order $\omega$), observe $(s, a_1, ..., a_{i-1})$
        \STATE Choose action $a_i \sim \pi_i(s, a_1, ..., a_{i-1})$
        \STATE Execute joint action $a = (a_1, ..., a_n)$
        \STATE Get observe $s'$, external reward $r^e$
        \STATE Store $(s, a, r^e, s')$ into buffer $\mathcal{D}_l$
        \IF{$t \% k == 0$}
            \STATE Compute upper-level reward $R_T = \sum_{t=0}^{k-1} r^e_t$
            \STATE Store $(s_T, \omega, R_T)$ into buffer $D_u$
        \ENDIF
    \ENDFOR
    \STATE Compute intrinsic reward $r^i_t = \frac{1}{k} A_\Omega(s_T, \omega)$
    \STATE Insert $r_t = r_{t}^{i}+r_{t}^{e}$ into $\mathcal{D}_l$ to replace $r^e$
    \STATE \textbf{Upper-policy improvement:}
    \STATE $\delta \gets r - Q_U(s, \omega, a)$
    \IF{$s'$ is non-terminal}
        \STATE $\delta \gets \delta + \gamma(1 - \beta_{\omega,\phi}(s')) Q_\Omega(s', \omega)$
        \STATE $\quad + \gamma \beta_{\omega,\phi}(s') \max_{\bar{\omega}} Q_\Omega(s', \bar{\omega})$
    \ENDIF
    \STATE $Q_U(s, \omega, a) \gets Q_U(s, \omega, a) + \alpha \delta$
    \STATE $v \gets v - \alpha_v \nabla_\phi \beta_{\omega,v}(s') (Q_\Omega(s', \omega) - \max Q_\Omega(s'))$
    \STATE \textbf{Lower-policy improvement:}
    \STATE Compute the advantage function $A_{\pi}$ for all $(s,a)$ to get $\mathcal{L}_{clip}=min(r_{\theta}A_{\pi},clip(r_{\theta}, 1 \pm \epsilon)A_{\pi})$
    \STATE Compute the policy entropy $S(\pi_{\theta}(s))$
    \STATE $\theta \gets \theta + \alpha_\theta \nabla_\theta \log \pi_{\omega,\theta}(a|s) [\mathcal{L}_{clip}+\eta S(\pi_{\theta}(s))]$
\ENDFOR
\end{algorithmic}
\end{algorithm}
Figure 3 illustrates our proposed method structure in detail. In our method, we add the option variable to the original STMG, i.e \(\Gamma =\left\langle\mathcal{I}, \mathcal{S},\left\{A^{i}\right\}_{i \in \mathcal{I}}, P,\left\{r^{i}\right\}_{i \in \mathcal{I}}, \gamma,\left\{h^{i}\right\}_{i \in \mathcal{I}},\{w\}_{w \in \Omega}\right\rangle\). For upper policy we draw on the framework of option-critic \cite{r14}, consisting of termination function and upper policy network.  Based on the current state of the system, the upper policy \(\pi_{w, \phi}(s)\) will choose the optimal execution order \(w\) for the state \(s\). Termination function \(\beta_{w,\varphi}(s)\) determines whether to halt the currently executing option based on the present state \(s\). Assuming that the number of agents in the system is \(n\), we abstract all the orders of the n agents as lower options, each order corresponding to an option. The option value function can be formalized as:
\begin{equation}
    Q_{\Omega}(s, w)=\sum_{a} \pi_{w, a}(a \mid s) Q_{U}(s, w, a)
\end{equation}
where \(Q_{U}:S \times \Omega \times A \rightarrow S^{\prime}\)is the augmented action value function:
\begin{equation}
    Q_{U}(s, w, a)=r(s, a)+\gamma \sum_{s^{\prime}} P\left(s^{\prime} \mid s, a\right) U\left(w, s^{\prime}\right)
\end{equation}
where \(U(s,w^{\prime})\) is the objective function of the termination function. It's to maximize the weighted sum expectation of the option value function and the termination state value function:
\begin{equation}
\begin{aligned}
        U\left(w, s^{\prime}\right)&=\left(1-\beta_{w, v}\left(s^{\prime}\right)\right) Q_{\Omega}\left(s^{\prime}, w\right)\\&+\beta_{w, \varphi}\left(s^{\prime}\right) V_{\Omega}\left(s^{\prime}\right)
\end{aligned}
\end{equation}
where \(\beta_{w,\varphi}(s^{\prime})\) is the probability value of terminating the current \(w\) given by the termination function based on \(s\), and \(V_{\Omega}(s^{\prime})\) is the state value function of terminating the current \(w\) obtained by the critic network. \(\phi\) and \(\varphi\) are the parameters of the critic network and termination function respectively.

For the lower policy, any algorithm with the Stackelberg game structure can be selected. We choose HAPPO and modify the input of its actor from the original \(s_{i}\) to \((s_{i},a_{1},...,a_{i-1})\) and use it as the lower policy of HPA.
\begin{figure}
    \centering
    \includegraphics[width=1\linewidth]{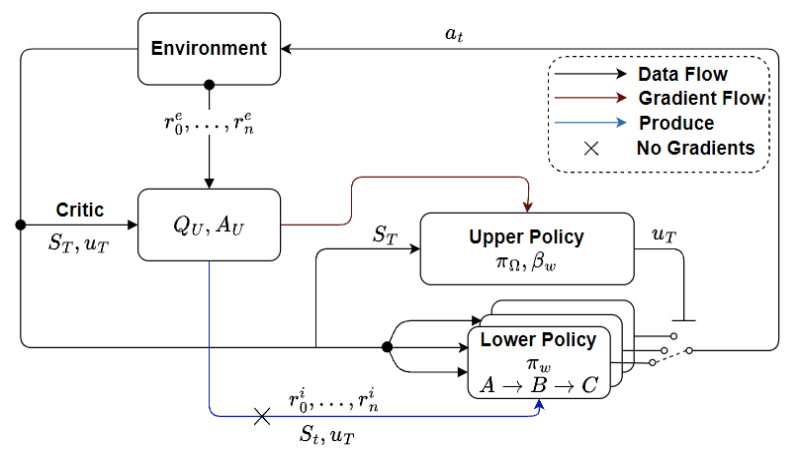}
    \caption{The overall architecture of HPA. }
    \label{fig:enter-label}
\end{figure}

\subsection{Hierarchical Priority Adjustment}
In the implementation of HPA: we disassemble the original option-critic framework, where the training between the upper policy and the lower policy is independent, which leads to the fact that there is no strong connection between the levels, and no communication is established between them. To address this problem we drew inspiration from HAVEN's \cite{r15} approach and divided HPA into two time scales, with the lower strategy being faster and the upper strategy being \(k\) times slower, and we denote the time scales of the upper and lower strategies by \(T\) and \(t\), respectively. Every \(k\) steps on the slow time scale \(\pi_{w}\) chooses \(w\), after selecting a particular lower strategy \(w\), the lower strategy \(\pi_{l}\) then selects an action \(a_{t}\) based on the local observation \(s_{i}\) of the \(k\) steps. All agents share a reward function given by the environment, which we denote \(r^{e}\) as external reward , and we also set the upper reward function to be shared, defined as \(R_{T}=\sum_{i=0}^{k-1}r^{e}\). We also denote the experience pools of different strategies by \(\mathcal{D}_{l}\) and \(\mathcal{D}_{u}\) to store \(\left\langle s_{t}, \mathbf{a}_{l}, r_{t}^{e}\right\rangle\) and \(\left\langle s_{T}, w_{T}, R_{T}\right\rangle\), respectively.

For the concurrent optimization of two-level strategies, the advantage function of \(\pi_{w}\) is used as the intrinsic reward \(r^{i}\), and the advantage function of the upper strategies can give the lower strategies a time abstraction of the next \(k\) steps to guide them to train. When \(\pi_{w}\) executing an action \(w_{T}\) in state \(s_{T}\), we set the advantage function of the upper policy as \(A_{h}(s_{T},w_{T})\), and then for \(\pi_{l}\), the advantage function is distributed over the \(k\) steps to obtain the intrinsic reward function of each underlying strategy, which can be expressed as :
\begin{equation}
    r_{t}^{i}=\frac{A_{h}\left(s_{T}, w_{T}\right)}{k}, T \cdot k \leq t \leq(T+1) \cdot k
\end{equation}
\(r^{i}\) and \(r^{e}\) as a joint reward function between all lower strategies, representing the coordination of strategies between levels and between agents respectively.

\subsection{Implementation}

\begin{figure*}[h]
    \centering
    \includegraphics[width=1\linewidth]{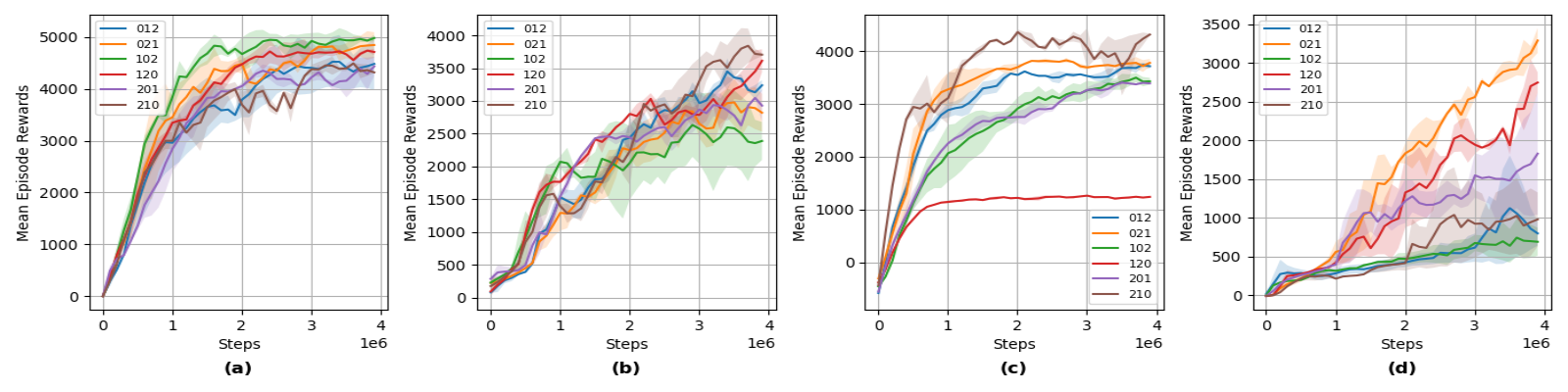}
    \caption{Experimental results with different execution sequences. Figure (a) and (b) show the experimental results of lower strategies with different execution orders on HalfCheetah \(3 \times 2\) and Walker2d \(3 \times 2\), respectively. Figure (c) and (d) show the different execution sequences of the STEP algorithm in these two environments.}
    \label{fig:enter-label}
\end{figure*}
\begin{figure*}[h]
    \centering
    \includegraphics[width=1\linewidth]{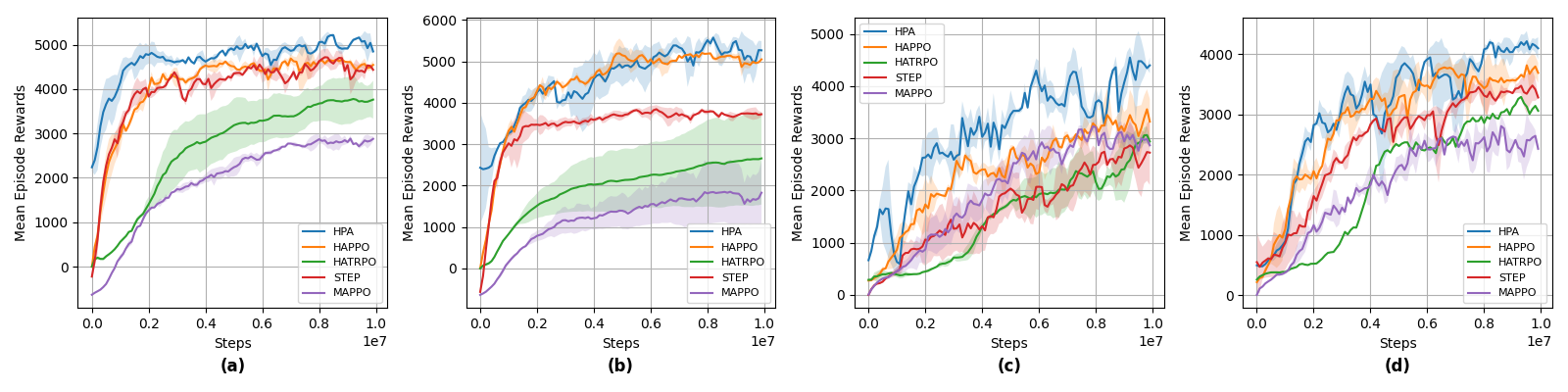}
    \caption{The mean episode rewards obtained by the proposed HPA and benchmarks.}
    \label{fig:enter-label}
\end{figure*}

\begin{figure*}
    \centering
    \begin{tabular}{c|c|c|c|c|c}
        \toprule
          & Rewards&  HalfCheetah 2x3& HalfCheetah 3x2 & Walker2d 2x3 & Walker2d 3x2 \\
        \midrule
        \multirow{2}{*}{HPA} 
          & mean &  \textbf{5176}&  \textbf{5345}&  \textbf{4532}& \textbf{4232}\\
        \cline{2-6}
          & max &  \textbf{5432}&  \textbf{5674}&  \textbf{5021}& \textbf{4467}\\
        \midrule
        \multirow{2}{*}{HAPPO} 
          & mean &  4502&  5100&  3461& 3785\\
        \cline{2-6}
          & max &  5002&  5419&  3957& 4001\\
        \midrule
        \multirow{2}{*}{HATRPO}
          & mean &  3853&  2727&  3089& 3142\\
        \cline{2-6}
          & max &  4238&  3800&  3090& 3285\\
        \midrule
        \multirow{2}{*}{MAPPO}
          & mean &  2980&  1903&  2971& 2673\\
        \cline{2-6}
          & max &  3002&  2654&  3499& 3000\\
        \midrule
        \multirow{2}{*}{STEP}
          & mean &  4394&  3898&  2845& 3451\\
        \cline{2-6}
          & max &  4899&  3902&  3291& 3788\\ 
        \bottomrule
    \end{tabular}
    \caption{Data characteristics. Mean denotes the mean episode rewards and max denotes the max episode rewards.}
    \label{tab:placeholder}
\end{figure*}
\begin{figure*}[h]
    \centering
    \includegraphics[width=1\linewidth]{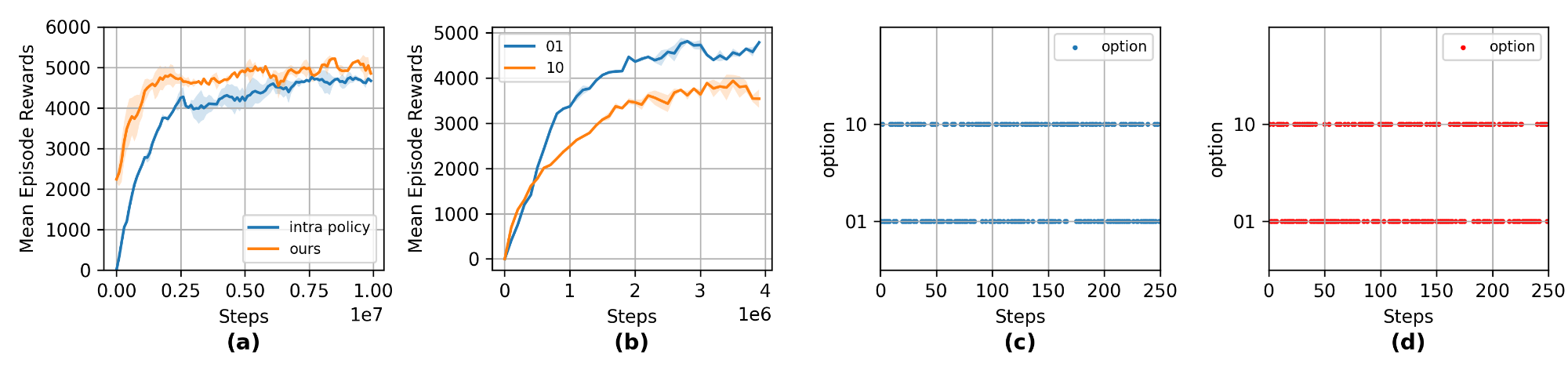}
    \caption{Visualisation experiments. (a) is the comparison experimental results of HPA and lower strategy. (b) is the experimental results showing different execution sequences of lower strategies. (c) and (d) are the visualisation experiments of sequential selection under random seeds of 1 and 10 after training.}
    \label{fig:enter-label}
\end{figure*}

We use the option-critic framework as our upper policy. At the same time, we chose the Proximal Policy Optimization (PPO) \cite{r17} algorithm for our implementation of option-critic. To ensure the balance between exploration and exploitation in the training phase, the upper options are selected by sampling actions from a Gaussian distribution generated by the strategy model. The loss function of the upper policy is:
\begin{equation}
\mathcal{L}(\theta)=\mathbb{E}_{s \sim P, w \sim \Omega}\left[\mathcal{L}_{\text {clip }}+\eta S\left(\pi_{\theta}(s)\right)\right] \\
\end{equation}\begin{equation}
    \mathcal{L}_{\text {clip }}=\min \left(r_{\theta} A_{\pi}, \operatorname{clip}\left(r_{\theta}, 1 \pm \varepsilon\right) A_{\pi}\right)
\end{equation}
where \(S(\cdot)\) is Shannon entropy used for strengthening exploration, \(r_{\theta}=\frac{\pi_{\theta}(w \mid s)}{\pi_{\theta_{\text {old }}}(w \mid s)}\) is the likelihood ratio between the current and previous policies, \(\epsilon\) is the clipping ratio, \(\eta\) is the coefficient of entropy. Critic network is used to fit the value function, and its loss function is expressed as: 
\begin{equation}
\begin{aligned}
\mathcal{L}(\phi) &= \max \Big[ \left(V_{\phi}(s_{T})-R_{T}\right)^{2}, \\
                  &\quad \quad\quad\left(\operatorname{clip}\left(V_{\phi}(s_{T}), V_{\phi_{dd}}(s_{T}) \pm \varepsilon\right)-R_{T}\right)^{2} \Big]
\end{aligned}
\end{equation}
where \(R_{T}\) is the cumulative value of the reward for the lower strategy executing \(k\) steps, and \(\epsilon\) is the clip coefficient. The loss function of the termination function is:
\begin{equation}
\begin{aligned}
\mathcal{L}(\varphi)
= \pi_{\Omega}\bigl(w_{T}\mid &s_{T}\bigr)\bigl[\,Q_{U}(s_{T},w_{T})\\
&\quad -\;\max_{u}Q_{U}(s_{T},w)+\psi\bigr]
\end{aligned}
\end{equation}
where \(\psi\) is the regularity coefficient. 

For the training of the lower strategy, we modify the reward of its loss function, replacing \(r^{e}\) with \(r=r^{i}+r^{e}\). The pseudo-code of the algorithm is shown in Algorithm 1.

\section{Experiments}

We evaluate the performance of our proposed algorithm, HPA. The main objectives of these experiments are: (a) verify that different orders of execution of the agents can affect the performance of the model; (b) evaluate our proposed method's performance in more challenging cooperative tasks; and (c) visualize that the model obtained from training selects the optimal order of the game in different situations. In the benchmark environment : the Multi-Agent MuJoCo \cite{r16}. Based on the reward settings (shared rewards) and the type of action control (continuous action space) in the environment, we compare HPA with various state-of-the-art MARL algorithms including algorithms based on the Stackelberg Game for facilitating multi-agent collaboration. 

\subsection{MuJoCo}
To measure the collaboration level of the agents, we choose representative cooperation scenarios such as Walker2d and HalfCheetah. We groupe the agents to reduce the exponential growth of the order of agents. For example, the agent is divided into 3 groups, each group contains two agents, which is denoted as \(3\times 2\). We denote the positions of the default order given by the environment as 0, 1 and 2, respectively, and all the orders are then denoted as full permutation of 0,1 and 2.

\subsection{Effect of  the method on different execution order}

In order to validate the theoretical proof process of question \textbf{(a)}, we use the advanced STEP algorithm and the HPA's lower policy algorithm to train all execution sequences of agents to compare the effect of different agents' execution sequences on the experimental results. We conducte the experiment using control tasks HalfCheetah and Walker2d provided by MuJoCo.

The results were shown in Figure 4. It can be found that the experimental results obtained by different orders of agents differ greatly, whether the current mature STEP algorithm or our proposed HPA's lower policy algorithm with the structure of the Stackelberg game. At the same time, this also triggers our thinking that when we use the Stackelberg game to conduct experiments, if we follow the environment's default order of agents for action input, there is a situation in which the order leads to suboptimal performance, rather than the upper limit of the model's performance. For example, in Fig. 4c, if the default order given by the environment is changed to 120, the performance of the STEP algorithm will be around 1200 rewards. But in fact, the effects of other orders are far higher than this result. 

\subsection{Comparison Results}

The proposed HPA method is compared with four state-of-the-art MARL algorithm: (1) MAPPO which is an extension of PPO based on a single agent to multi-agent system. (2) HAPPO and (3) HATRPO which are representative examples of heterogeneous agent algorithms, with strong theoretical support and performance guarantees. (4) STEP which is currently the most advanced algorithm for modelling multi-agent systems using Stackelberg games. 

The comparison results are shown in Figure 5 and Figure 6. HPA performed better than other algorithms in four experiments. Specifically, in the environment of HalfCheetah \(2 \times 3\), the proposed method performed the best among all methods and gained nearly 5200 rewards, while the second-best method was HAPPO method that gained nearly 4500 rewards. In the environment of  HalfCheetah \(3 \times 2\), the proposed method and HAPPO performed the best among all methods and gained nearly 5500 rewards, which may be because the agent partitioning method is insensitive to the priority of execution. Compared to the HalfCheetah environment, the Walker2d environment places higher demands on collaboration between agents: the goal of HalfCheetah is for agents to move as quickly as possible by coordinating their joints, while Walker2d not only requires movement, but also requires agents to maintain balance. In the environment of Walker2d, the performance improvement of our proposed method is much higher than that of HalfCheetah. This is because when it is necessary to maintain the balance of the agent body, the HPA algorithm can select the optimal execution sequence based on the current posture of the intelligent body to adjust its posture and achieve balance. The comparison results show that the proposed method achieves the best performance in four complex cooperative environments.

\subsection{Ablation Studies}

This section aims to verify question \textbf{(b)} whether the upper-level sequence adjustment strategy of HPA is truly effective. We selecte the HalfCheetah \(2 \times 3\) experimental environment, and the experimental results are shown in Figure 7. Fig. 7a shows the experimental results comparing the HPA algorithm with the default sequence of the lower strategy. It can be seen that HPA is significantly better than the lower strategy. By observing Fig. 7c and Fig. 7d, we can see that during execution, the upper policy selects different execution sequences based on the changes of the system state. Execution sequences 01 and 10 each account for approximately \(50\%\). As shown in Fig. 7b, even though sequence 10 is less effective than 01, the upper strategy still chooses 10 under certain state. This indicates that during movement, different agent's postures need to be adjusted by different execution sequences in order to achieve long-term goals. This also confirms that the necessity of sequence adjustment in N-level Stackelberg Game.

\begin{figure}
    \centering
    \includegraphics[width=1\linewidth]{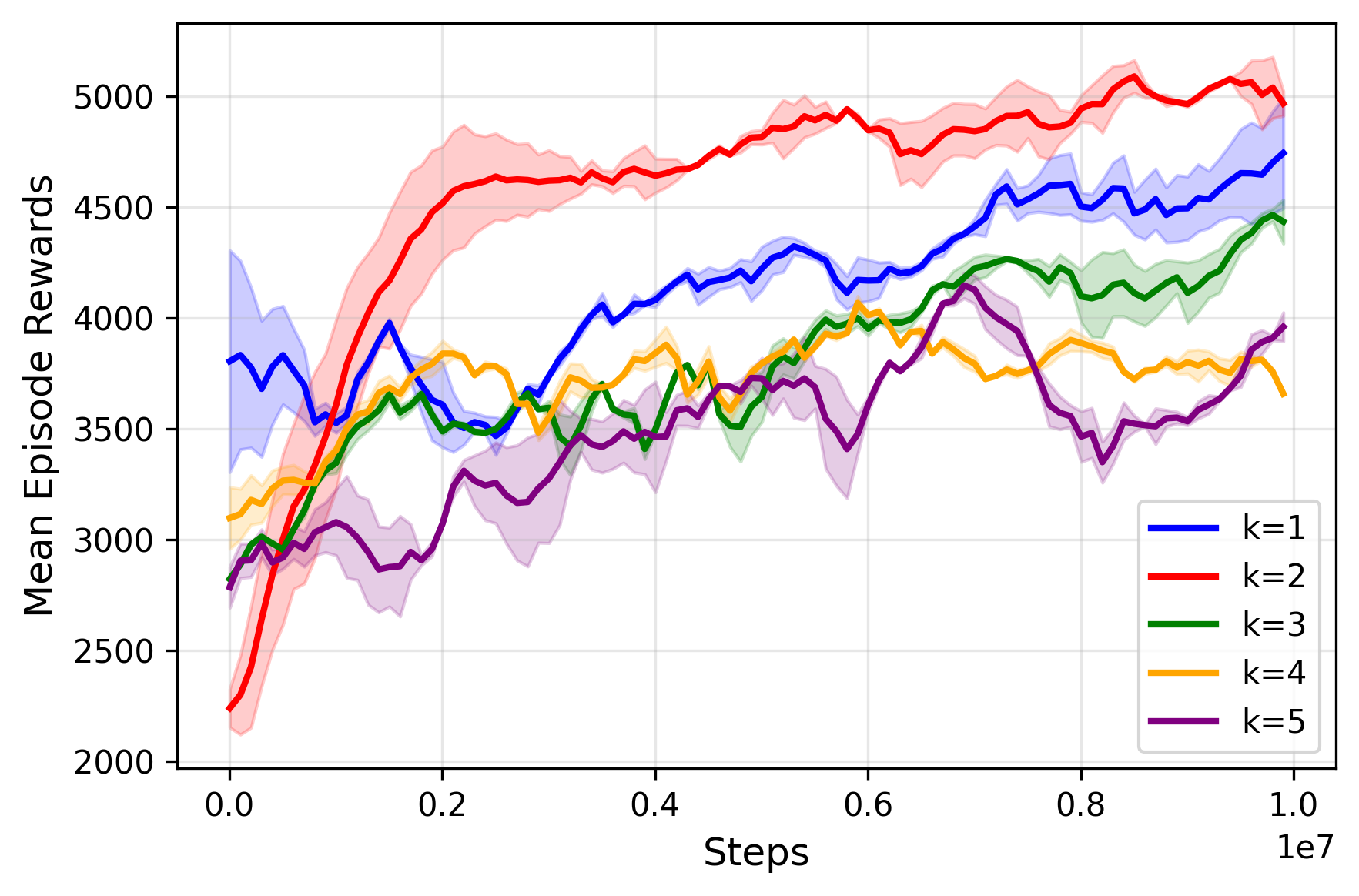}
    \caption{Influence of the k for HAP}
    \label{fig:placeholder}
\end{figure}

We demonstrate the impact of hyperparameter k settings on HAP performance in the HalfCheetah \(2\times3\) in Figure 8. We discuss how the k influences the performance. First, when k equals 1, the fast-slow update strategy degenerates into a scenario where only the upper-level strategy provides intrinsic rewards to the lower-level strategy, with no multi-step guidance. Consequently, its performance deteriorates relative to that observed when k equals 2. In addition, the results show a trend that HAP performs worse as the k increases. This is because when k becomes large, if the agent's posture requires adjustment during execution, it cannot promptly make adjustments. The value of the hyperparameter k should not be excessively large.

\section{Conclusion}

In this paper, we solve question \textbf{(a)} via a game-theoretic formulation and optimality conditions. It has been proven that, in general, changes in the execution order of agents will lead to shifts in SE points. To solve the question \textbf{(b)}, we propose the HPA framework, which employs a HRL scheme: an upper policy dynamically selects the optimal agent ordering based on the current state, while lower-level agents execute actions in STMG according to the chosen order. Finally, the experiments on multi-agent MuJoCo control tasks demonstrate that HPA outperforms state-of-the-art baselines, validating both the theoretical proof and the practical effectiveness of dynamic priority scheduling method. The shortcoming of our model lies in the fact that while multi-agent grouping mitigates the exponential growth of sequential permutations, it simultaneously constrains the potential performance of certain sequences. Moreover, the partitioning method for multi-agent components also influences game outcomes. Future work will address the limitations of sequence quantities by exploring more adaptive dynamic adjustment mechanisms for the execution order of agents, thereby enhancing the overall system efficiency and robustness in diverse scenarios. 

\newpage
\bibliographystyle{ACM-Reference-Format}
\bibliography{refs}

@article{r1,
  title={Nash Q-learning for general-sum stochastic games},
  author={Hu, Junling and Wellman, Michael P},
  journal={Journal of machine learning research},
  volume={4},
  number={Nov},
  pages={1039--1069},
  year={2003}
}

@inproceedings{r2,
  title={Mean field multi-agent reinforcement learning},
  author={Yang, Yaodong and Luo, Rui and Li, Minne and Zhou, Ming and Zhang, Weinan and Wang, Jun},
  booktitle={International conference on machine learning},
  pages={5571--5580},
  year={2018},
  organization={PMLR}
}

@inproceedings{r3,
title={Trust Region Policy Optimisation in Multi-Agent Reinforcement Learning},
author={Jakub Grudzien Kuba and Ruiqing Chen and Muning Wen and Ying Wen and Fanglei Sun and Jun Wang and Yaodong Yang},
booktitle={International Conference on Learning Representations},
year={2022},
url={https://openreview.net/forum?id=EcGGFkNTxdJ}
}

@book{r4,
  title={Market structure and equilibrium},
  author={Von Stackelberg, Heinrich},
  year={2010},
  publisher={Springer Science \& Business Media}
}

@inproceedings{r5,
  title={Bi-level actor-critic for multi-agent coordination},
  author={Zhang, Haifeng and Chen, Weizhe and Huang, Zeren and Li, Minne and Yang, Yaodong and Zhang, Weinan and Wang, Jun},
  booktitle={Proceedings of the AAAI conference on artificial intelligence},
  volume={34},
  number={05},
  pages={7325--7332},
  year={2020}
}

@INPROCEEDINGS{r6,
  author={Yang, Boling and Zheng, Liyuan and Ratliff, Lillian J. and Boots, Byron and Smith, Joshua R.},
  booktitle={2023 IEEE International Conference on Robotics and Automation (ICRA)}, 
  title={Stackelberg Games for Learning Emergent Behaviors During Competitive Autocurricula}, 
  year={2023},
  volume={},
  number={},
  pages={5501-5507},
  doi={10.1109/ICRA48891.2023.10160875}}

@inproceedings{r7,
  title     = {Inducing Stackelberg Equilibrium through Spatio-Temporal Sequential Decision-Making in Multi-Agent Reinforcement Learning},
  author    = {Zhang, Bin and Li, Lijuan and Xu, Zhiwei and Li, Dapeng and Fan, Guoliang},
  booktitle = {Proceedings of the Thirty-Second International Joint Conference on
               Artificial Intelligence, {IJCAI-23}},
  publisher = {International Joint Conferences on Artificial Intelligence Organization},
  editor    = {Edith Elkind},
  pages     = {353--361},
  year      = {2023},
  month     = {8},
  note      = {Main Track},
  doi       = {10.24963/ijcai.2023/40},
  url       = {https://doi.org/10.24963/ijcai.2023/40},
}

@InProceedings{r9,
  title = 	 {Sequential Asynchronous Action Coordination in Multi-Agent Systems: A Stackelberg Decision Transformer Approach},
  author =       {Zhang, Bin and Mao, Hangyu and Li, Lijuan and Xu, Zhiwei and Li, Dapeng and Zhao, Rui and Fan, Guoliang},
  booktitle = 	 {Proceedings of the 41st International Conference on Machine Learning},
  pages = 	 {59559--59575},
  year = 	 {2024},
  editor = 	 {Salakhutdinov, Ruslan and Kolter, Zico and Heller, Katherine and Weller, Adrian and Oliver, Nuria and Scarlett, Jonathan and Berkenkamp, Felix},
  volume = 	 {235},
  series = 	 {Proceedings of Machine Learning Research},
  month = 	 {21--27 Jul},
  publisher =    {PMLR},
  url = 	 {https://proceedings.mlr.press/v235/zhang24au.html},
}

@inproceedings{r10,
  title={Who Plays First? Optimizing the Order of Play in Stackelberg Games with Many Robots},
  author={Hu, Haimin and Dragotto, Gabriele and Zhang, Zixu and Liang, Kaiqu and Stellato, Bartolomeo and Fisac, Jaime F},
  booktitle={Proceedings of Robotics: Science and Systems},
  year={2024}
}

@article{r11,
  title={Accurate solution to overdetermined linear equations with errors using L1 norm minimization},
  author={Rosen, J Ben and Park, Haesun and Glick, John and Zhang, Lei},
  journal={Computational optimization and applications},
  volume={17},
  number={2},
  pages={329--341},
  year={2000},
  publisher={Springer}
}

@inproceedings{r12,
  title={The Levenberg-Marquardt algorithm: implementation and theory},
  author={Mor{\'e}, Jorge J},
  booktitle={Numerical analysis: proceedings of the biennial Conference held at Dundee, June 28--July 1, 1977},
  pages={105--116},
  year={2006},
  organization={Springer}
}

@article{r13,
  title={Computing a trust region step},
  author={Mor{\'e}, Jorge J and Sorensen, Danny C},
  journal={SIAM Journal on scientific and statistical computing},
  volume={4},
  number={3},
  pages={553--572},
  year={1983},
  publisher={SIAM}
}

@inproceedings{r14,
  title={The option-critic architecture},
  author={Bacon, Pierre-Luc and Harb, Jean and Precup, Doina},
  booktitle={Proceedings of the AAAI conference on artificial intelligence},
  volume={31},
  number={1},
  year={2017}
}

@inproceedings{r15,
  title={Haven: Hierarchical cooperative multi-agent reinforcement learning with dual coordination mechanism},
  author={Xu, Zhiwei and Bai, Yunpeng and Zhang, Bin and Li, Dapeng and Fan, Guoliang},
  booktitle={Proceedings of the AAAI conference on artificial intelligence},
  volume={37},
  number={10},
  pages={11735--11743},
  year={2023}
}

@INPROCEEDINGS{r16,
  author={Todorov, Emanuel and Erez, Tom and Tassa, Yuval},
  booktitle={2012 IEEE/RSJ International Conference on Intelligent Robots and Systems}, 
  title={MuJoCo: A physics engine for model-based control}, 
  year={2012},
  volume={},
  number={},
  pages={5026-5033},
  doi={10.1109/IROS.2012.6386109}}

@article{r17,
  title={Proximal policy optimization algorithms},
  author={Schulman, John and Wolski, Filip and Dhariwal, Prafulla and Radford, Alec and Klimov, Oleg},
  journal={arXiv preprint arXiv:1707.06347},
  year={2017}
}

@incollection{r19,
  title={Reexamination of the perfectness concept for equilibrium points in extensive games},
  author={Bielefeld, R Selten},
  booktitle={Models of strategic rationality},
  pages={1--31},
  year={1988},
  publisher={Springer}
}


\end{document}